# Large-scale epitaxy of two-dimensional van der Waals room-temperature ferromagnet Fe$_5$GeTe$_2$

**Authors**

Mário Ribeiro[1*], Giulio Gentile[1], Alain Marty[1], Djordje Dosenovic[2], Hanako Okuno[2], Céline Vergnaud[1], Jean-François Jacquot[3], Denis Jalabert[2], Danilo Longo[4], Philippe Ohresser[4], Ali Hallal[1], Mairbek Chshiev[1,5], Olivier Boulle[1], Frédéric Bonell[1*], Matthieu Jamet[1]

[1] University Grenoble Alpes, CEA, CNRS, IRIG-Spintec, F-38000 Grenoble, France

[2] University Grenoble Alpes, CEA, IRIG-MEM, F-38000 Grenoble, France

[3] University Grenoble Alpes, CEA, IRIG-SYMMES, F-38000 Grenoble, France

[4] Synchrotron SOLEIL, L'Orme des Merisiers, Saint-Aubin, 91192 Gif-sur-Yvette, France

[5] Institut Universitaire de France (IUF), 75231, Paris, France

* Corresponding authors: mario.oliveiraribeiro@cea.fr; frederic.bonell@cea.fr

**Abstract**

In recent years, two-dimensional van der Waals materials have emerged as an important platform for the observation of long-range ferromagnetic order in atomically thin layers. Although heterostructures of such materials can be conceived to harness and couple a wide range of magneto-optical and magneto-electrical properties, technologically relevant applications require Curie temperatures at or above room-temperature and the ability to grow films over large areas. Here we demonstrate the large-area growth of single-crystal ultrathin films of stoichiometric $Fe_5GeTe_2$ on an insulating substrate using molecular beam epitaxy. Magnetic measurements show the persistence of soft ferromagnetism up to room temperature, with a Curie temperature of 293 K, and a weak out-of-plane magnetocrystalline anisotropy. Surface, chemical, and structural characterizations confirm the layer-by-layer growth, 5:1:2 Fe:Ge:Te stoichiometric elementary composition, and single crystalline character of the films.

**MAIN TEXT**

**Introduction**

Until recently, long-range magnetic ordering was not thought possible in two-dimensional (2D) systems. The Mermin-Wagner theorem states that thermal fluctuations in a low-dimensionality isotropic Heisenberg model block the emergence of ferromagnetism (FM)[1]. In 2017, ferromagnetism was observed in exfoliated flakes of the 2D van der Waals (vdW) material $Cr_2Ge_2Te_6$ [2], and later in $CrI_3$[3]. These materials are electrical insulators, exhibit perpendicular magnetic anisotropy, and have Curie temperature ($T_C$) well below room temperature (RT). The prevailing understanding is that long-range ferromagnetic order in 2D vdW systems survives due to the spin-wave excitation gap opened by the strong intrinsic magnetocrystalline anisotropy (MCA). 2D vdW materials became a rich platform to explore low-dimensionality magnetism, and in a few years the list of 2D vdW FMs increased with the addition of $FePS_3$[4], $Fe_3GeTe_2$[5], $Fe_4GeTe_2$[6], and $Fe_{5-\delta}GeTe_2$[7]. Fe-Ge-Te (FGT) ternary compounds in particular stood out due to their itinerant ferromagnetism, which is of great interest for spintronic applications.

Among the family of 2D vdW FMs, FGT compounds host a variety of noticeable magnetic effects, namely large anomalous- and planar topological Hall effect[8,9], magnetic skyrmions[10,11], Kondo lattices[12], and giant tunneling magnetoresistance[13]. To the best of our knowledge, the highest $T_C$ in pristine, bulk 2D vdW layered materials has been observed in $Fe_{5-\delta}GeTe_2$, with $T_C \sim 310$ K[7]. $Fe_3GeTe_2$ exhibits a $T_C$ of 230 K[14,15] in bulk form and 130 K[5] for a single layer. In these systems, the $T_C$ can be tuned via the use of extrinsic mechanisms such as ionic gating[16], doping[17], ion implantation[18], proximity to topological insulators[19], and patterning methods[20]. Metastable states can be stabilized via temperature quenching and temperature-cycles[21]. From a technological standpoint, high $T_C$ is among the magnetic properties of highest interest for next-generation spintronic applications integrating 2D vdW FMs. $Fe_{5-\delta}GeTe_2$ stands out for its near RT $T_C$ and itinerant ferromagnetism. However, the mechanisms governing its magnetic properties are still unclear, with variable magnetic properties ($T_C$ and anisotropy) being reported. Most experimental realizations have been done using Fe-deficient bulk crystals or micrometer-sized flakes exfoliated from them [7,21–23]. The properties of $Fe_5GeTe_2$ films with exact 5:1:2 stoichiometry remain unexplored. Recent studies pointed out the role played by Fe vacancies and the different Fe sublattices on the complex range of magnetic behaviors observed[6,7,21]. Additionally, magnetic critical exponents of bulk $Fe_{5-\delta}GeTe_2$ are close to those of a 3D Heisenberg and 3D XY systems[23], remaining unclear if these models still hold in ultrathin regimes.

While many of the demonstrations of vdW FMs materials relied on bulk crystals and on exfoliated flakes in the 2D limit, growth by molecular beam epitaxy (MBE) has been pursued to expand the complexity of the designed heterostructures and large-area scalability[19]. MBE is a well-established deposition technique for the realization of atomically thin single-crystal vdW materials and heterostructures with sharp interfaces[24]. Any outlook on future heterostructures and device concepts based on 2D vdW FMs requires the large-area fabrication of high-quality ultrathin films[24–28], for which MBE is well suited.

Here, we demonstrate the MBE growth of single-crystal thin films of stoichiometric $Fe_5GeTe_2$ on an insulating substrate, $Al_2O_3$(001). In-situ reflection high-energy electron diffraction (RHEED) measurements demonstrate the layer-by-layer growth of single-crystalline films. Chemical and structural characterization using Rutherford backscattering spectroscopy (RBS), scanning transmission electron microscopy (STEM), and X-ray diffraction (XRD), confirm the 5:1:2 Fe:Ge:Te stoichiometry and rhombohedral crystal structure. Superconducting quantum interference device (SQUID) and X-ray magnetic circular dichroism measurements (XMCD) show a large magnetization saturation ($M_S$) of 644 kA/m at 10 K (1.95 $\mu_B$/Fe), $T_C$ of 293 K, and a magnetic critical exponent $\beta$ of 0.293, which approaches the universal window for critical-exponents in 2D systems.

**Results**

The MBE growth of $Fe_5GeTe_2$ was carried out in a home-built UHV deposition system with a base pressure in the low $10^{-10}$ mbar (see Methods for further information). The $Al_2O_3$(001) substrates were first annealed in air at 1000 ºC in a tube furnace to promote the formation of a terrace-and-step smooth surface, and again annealed in the UHV chamber at 800 ºC for 20 minutes prior to the deposition. Ge and Te were co-deposited using Knudsen effusion and thermal cracker cells, respectively, and Fe using e-beam evaporation. The substrate temperature for the growth of $Fe_5GeTe_2$ was set to 350 ºC, the flux ratio of Fe:Ge was kept to the stoichiometric 5:1, and the Te flux on overpressure (approximately twice the stoichiometric flux). After the deposition, the films were annealed for 20 minutes at 550 ºC under Te flux, and then capped with a 3 nm-thick Al film at RT. The growth of $Fe_5GeTe_2$ on $Al_2O_3(001)$ was monitored *in-situ* using RHEED. We report mainly on the characterization of 12-nm-thick $Fe_5GeTe_2$ films (1 monolayer ~ 0.977 nm). In Figure 1A, the top left side panel shows well-defined RHEED patterns of the $Al_2O_3(001)$ surface with the electron-beam (e-beam) aligned along [110], while the bottom left panel shows the diffraction pattern for the same direction after the growth of 12 nm of $Fe_5GeTe_2$. Similarly, the top and bottom right-side panels show the diffraction patterns of the $Al_2O_3(001)$ surface before and after the growth of 12 $Fe_5GeTe_2$ layers, respectively, with the e-beam aligned along [210] (see Supplementary Figure 1 for RHEED snapshots during the deposition and annealing of 12-nm and 2-nm-thick samples).

Despite the large lattice misfit with $Al_2O_3(001)$(~-15%), the sharp streaks and anisotropic RHEED patterns demonstrate the single crystalline character of $Fe_5GeTe_2$, with *c*-axis perpendicular to the surface. The two azimuths are aligned with those of the $Al_2O_3(001)$ surface, showing the $Al_2O_3(001)[100]/Fe_5GeTe_2(001)[100]$ epitaxial relationship. The real-time monitoring of the RHEED intensity of the 1$^{st}$ order diffraction streaks allowed for the observation of periodical intensity oscillations characteristic of a layer-by-layer growth mode (Figure 1B). By indexing the successive oscillations as the formation of a single layer, the rate of growth was estimated at 1.69 monolayers per minute of deposition. Atomic force microscopy (AFM) scans of the surface of 12-nm-thick films not capped with Al, taken within minutes of exposing the surface to the environment, show a smooth surface with root-mean-square roughness (RMS) of 0.5 nm, without clearly defined terraces, but with atomic steps corresponding to those of one or multiple monolayers (Figure 1C).

To gather further information on the crystallographic structure, domain size, mosaic spread, and lattice parameter distribution of the films deposited, we performed in-plane (grazing incidence) and out-of-plane XRD measurements (see Methods for further details). $Fe_5GeTe_2$ is a highly anisotropic layered vdW ferromagnet with trigonal crystal system. Typical lattice constants reported in literature for bulk $Fe_{5-\delta}GeTe_2$ are $a \approx 4.04$ Å and $c \approx 29.19$ Å[7,21,29]. Unit cells are typically represented with three stacked $Fe_5GeTe_2$ layers, with vdW gaps between adjacent Te planes. The complex atomic arrangement, the common presence of stacking faults and vacancy-induced disorder makes the precise determination of the crystallographic structure an open issue[7,21,22,29]. In Figure 2A, we depict the unit cell and three $Fe_5GeTe_2$ layers viewed along the [110] direction, according to the crystallographic model of ref. [22]. The model considers Fe-deficiency via fractional occupation probabilities of the Fe sites, and includes a split Ge and Fe1 site with half-occupation probability. Figure 2B shows the out-of-plane $\theta/2\theta$ diffraction scan of a 12-nm-thick film, where we index the $Al_2O_3(00l)$ and $Fe_5GeTe_2(00l)$ Bragg peaks. The determined unit cell lattice constant $c \approx 29.3$ Å is in good agreement with previous reports (thickness of 1 ML = $c/3$)[7,21,29]. The fringes around the Bragg peaks of $Fe_5GeTe_2(009)$ indicate low surface and interface roughness (also confirmed by the RMS of 0.5 nm measured by AFM in Figure 1C), and allows for the estimation of a film thickness of $12.5 \pm 0.3$ nm, in strong agreement with the expected growth of 12 layers determined from the RHEED oscillations (See Supplementary Figure 2 for the XRD analysis of a 2-nm-thick film). In Figure 2C, we show in-plane radial scans $\theta_{//}/2\theta_{//}$ that were acquired along the ($hh0$) and ($0h0$) reciprocal directions, 30º apart from each other. The full width at half maximum (FWHM) is also displayed above the peaks. From these scans, we obtain an in-plane lattice parameter $a = 4.068$ Å, expanded by roughly 0.7% compared to the reference value of 4.04 Å. The mutually exclusive presence of the ($hh0$) and ($0h0$) peaks on each radial scan indicate the absence of domains rotated by at least 30º. This is further confirmed in Figure 2D, which shows the azimuthal dependence of the intensity of the Bragg peaks identified in the radial scans (grey curves). The baseline of the azimuthal scans is the sum of the instrumental background and of the signal from misaligned crystallites with isotropic distribution of orientations. The observed correspondence between the azimuthal and radial base lines implies that there is no isotropic component in the crystallites angular distribution. The film is single crystalline with the expected 60º periodicity of the basal plane of the crystal lattice. Noticeably, the FWHM for the Bragg peaks is substantially larger. From the study of the momentum transfer for each radial and azimuthal Bragg diffraction peak and its dispersion, one can estimate the domain size $D$, lattice parameter distribution $\Delta a/a$, and in-plane mosaic spread $\Delta \xi$ of the film[31,32] (see Supplementary Note 1 and Supplementary Figure 3 for details on the calculations). For the 12-nm-thick film, $D$ was determined to be 33 nm, which must be considered as a lower bound for the domain size due to the instrumental limited resolution, $\Delta a/a \approx 0.66\%$, and $\Delta \xi \approx 3.34º$, values that corroborate the single-crystalline character of the deposited films.

In order to take a closer look at disorder and stacking of the $Fe_5GeTe_2$ ultrathin films, we recorded atomic-resolution real-space images by STEM (see Methods for further information). Figure 3A shows a high-angle annular dark-field (HAADF) cross-section image along [110]. The atomically flat $Al_2O_3$ promoted the epitaxy of well-oriented layers that are uniform over large areas, parallel to each other, and with no perceivable rotated domains. At the interface between the two materials, there is a thin amorphized gap of 6 Å, after which the first $Fe_5GeTe_2$ layer develops.

Chemical analysis using RBS spectroscopy confirms the nominal 5:1:2 stoichiometry of the Fe:Ge:Te elementary composition of the film (inset of Figure 3A). Energy dispersive X-ray spectroscopy (EDX) shows a homogenous distribution of each element in the film (Figure 3B). In Figure 3C, $Z$ contrast enhanced HAADF over a smaller section of the film yields a site distribution with clear indication of the presence of heavier atoms at the edges of the layers. Element resolved EDX over two layers shows the clear presence of Fe and Ge in between Te sites. Fe shows a broader dependence on position than Ge, as expected from its distribution in the cell. The presence of the Fe1 atom is not clearly resolved in our measurements, but the symmetry of the images points to a 50% occupation probability in the two possible sites, as in the model of ref. [22]. In order to confirm it, we performed simulations of Z-contrast HAADF images for several occupation probabilities (Figure 3D). For a 100% occupation probability, the simulation indicates that an STEM scan should be able to clearly resolve the site occupied by the Fe1 atom, while at 0% the background strongly contrasts with the atomic sites. At 50%, we observe a tail connecting the inner Fe2 atoms to Te. This reproduces well the experimental result, and strongly suggests that a crystallographic model with a Fe1 half-split site applies to our epitaxial films.

The magnetic properties of the films were characterized using SQUID and XMCD (see Materials and Methods for further information). In Figure 4A and Figure 4B we plot isothermal magnetic hysteresis loops at different temperatures with the magnetic field applied in the *ab* basal plane, and along the *c*-axis of the layers, respectively. The data have been corrected by subtracting the magnetic background observed at temperatures well above $T_C$, and a residual slope due to a paramagnetic contribution from the substrate below 50 K. The same slope was subtracted from the in-plane and out-of-plane curves. This treatment assumes that the high temperature magnetic signal originates from the substrate, which is justified by our XMCD and SQUID measurements, as explained in the following. The insets show the raw magnetization curves and the considered background in grey. The in-plane hysteresis loops show low coercivity (~11 mT) and saturation fields (~200 mT). The saturation magnetization averaged over in-plane and out-of-plane measurements amounts to $M_s = (644 \pm 54)$ kA/m at 10 K, corresponding to a magnetic moment $m = (1.95 \pm 0.12)$ $\mu_B$/Fe. These values are in-line with those previously reported for Fe-deficient $Fe_{5-\delta}GeTe_2$ ($\delta = 0.1 - 0.3$) films exfoliated from bulk crystals [21,33]. Perpendicular to the surface, the saturation field largely increases, indicating an easy-plane magnetic anisotropy.

In Figure 4C, the temperature dependence of the magnetization remanence ($M_r$) closely follows the scaling behavior of an ordered ferromagnetic system, $M_r \propto \left(\frac{T_C-T}{T_C}\right)^\beta$, yielding $T_C \approx 293$ K and $\beta \approx 0.294$. This behavior is in striking contrast to the complex temperature dependences observed in bulk and Fe-deficient $Fe_{5-\delta}GeTe_2$ films[7,21], suggesting that MBE-grown films present a higher structural and chemical homogeneity. The $M_r$ can be reasonably fitted with 3D models (see Supplementary Figure 4). However, our determined $\beta$ significantly deviates from the corresponding magnetic critical exponent and from the one observed in bulk-crystals $\beta \approx 0.346$[23]. With a value of $\beta \approx 0.294$, our results approaches the window for critical exponents of two-dimensional systems[34], and possibly indicates a characteristic 2D behavior in ultrathin films. Figure 4D shows the normalized in-plane and out-of-plane isothermal magnetic loops at 50 K, 100 K, 200 K, and 300 K. We determine the effective anisotropy $K_{eff}$ from the area enclosed between the curves. The uniaxial anisotropy $K_u$ is deduced from $K_{eff}$ and the measured $M_S$, with the relation $K_{eff} = K_u + K_d = K_u - (\mu_0/2)M_S^2$, where $K_d$ is the shape anisotropy and $\mu_0$ the vacuum permeability. The temperature dependence of $K_{eff}$, $K_u$, and $K_d$ is displayed on Figure 4E. We estimate a small positive uniaxial anisotropy $K_u \approx 0.09$ J/cm$^3$ that favors an easy $c$-axis, but remains weaker than shape anisotropy at all temperatures. Figure 4F shows the magnetic moments of this sample (S1) determined by SQUID and compares it to the magnetic moment of a second sample (S2) measured by both SQUID and XMCD. Sample S2 was capped with 5 nm of Te instead of Al. For sample S2, the excellent agreement between SQUID and XMCD confirms the validity of the background correction of SQUID measurements. The smaller magnetic moment of sample S2 compared to that of sample S1 (whose 5:1:2 Fe:Ge:Te composition was ascertained by RBS, in Figure 3A) might originate from Fe deficiency. In Figure 4G,H we detail the XMCD characterization of sample S2, and show the XMCD spectra at the Fe and Ge L$_{2,3}$ edges at 4 K (see Methods for further details). The measurements were performed in total electron yield mode, with X-ray incidence normal to the surface ($\alpha = 90°$), and the ±5 T magnetic field parallel to the beam direction. Additional measurements were carried out at $\alpha = 30°$ (not shown). We applied the conventional sum rules analysis[35,36] to the Fe absorption signal in order to determine the effective spin ($m_S^{eff,\alpha} = m_S + m_D^\alpha$) and orbital ($m_L^\alpha$) moments (see Supplementary Note 2 for further details on the determination of the Fe magnetic moments with XMCD sum rules). From those, we derive the isotropic spin ($m_S$) and orbital ($m_L$) moments, as well as the intra-atomic dipole moment anisotropy $\Delta m_D = m_D^\perp - m_D^\parallel$ and the orbital moment anisotropy $\Delta m_L = m_L^\perp - m_L^\parallel$ ($\perp$ and $\parallel$ denote the out-of-plane and in-plane components, respectively). We obtained $m_S = 1.54$ µ$_B$, $m_L = 0.08$ µ$_B$, $\Delta m_D = 0.21$ µ$_B$ and $\Delta m_L = -0.01$ µ$_B$. The total magnetic moment amounts to $m_S + m_L = 1.62$ µ$_B$, in excellent agreement with the theoretical value determined from first principles reported in Table 1 (see Methods for further details).

The small $m_L$ and $\Delta m_L$ are consistent with the weak out-of-plane uniaxial magnetic anisotropy deduced by SQUID measurements. We also observed a small polarization of Ge atoms, in agreement with ref. [33] and with our *ab initio* results (Table 1). As depicted in Figure 4H, both 2$p$→4$d$ ($\Delta L$=+1) and 2$p$→4$s$ ($\Delta L$=-1) transitions present a finite XMCD signal. The sign of the XMCD for 4$d$ (4$s$) final states is equal (opposite) to the one of Fe. Given the opposite sign of $\Delta L$ for these transitions, we deduce that the magnetic moment of both 4$d$ and 4$s$ bands is opposite to the Fe one. In ref. [33], a sizeable magnetic moment was also observed on Te atoms, but we could not confirm it in our film due to the use of a Te capping layer. Note that we neglected the Ge and Te contributions to determine the Fe magnetic moment by SQUID. According to Table 1, this assumptions leads to an underestimation of the Fe magnetic moment by only 2%.

**Discussion**

In summary, we demonstrated the large-scale growth of few-layer, single-crystal $Fe_5GeTe_2$ on $Al_2O_3(001)$ using molecular beam epitaxy, and provided a thorough structural and magnetic characterization. We showed that the MBE technique yields $Fe_5GeTe_2$ layers with high structural quality, layer-by-layer growth, virtually no Fe deficiency, and room temperature ferromagnetism. The ultrathin films exhibit soft ferromagnetism, a large magnetization of about 644 kA/m at 10 K, easy-plane magnetization, with a weak $c$-axis magnetocrystalline anisotropy $K_u \sim 0.09$ J/cm$^3$. The deviation of the magnetic critical behavior from a 3D Heisenberg/3D XY model to one close to that of 2D systems, with $\beta = 0.294$ might be a fingerprint of the effects of reduced dimensionality on the physical properties of the films. STEM images and simulations clearly indicate a crystallographic structure with half-split occupation probability of the Fe1 atomic site.

Our work highlights the role of MBE as a key tool for the fine tailoring of 2D vdW FMs single crystals. In the particular case of $Fe_5GeTe_2$, we have grown a benchmark material with exact stoichiometry and well-defined crystalline structure to well-establish the magnetic properties of this novel vdW ferromagnetic material. It will serve as a platform to further unveil the complex dependence of the magnetic properties on the film composition and crystal structure, and as a stepping stone towards the integration in large-scale vdW multilayers for spintronics.

**Methods**

**MBE**

$1\times 1$ cm$^2$ $Al_2O_3(001)$ substrates were cleaned prior to the tube furnace annealing with a standard isopropanol/acetone ultrasonic bath. For the growth of $Fe_5GeTe_2$, high purity Ge (99.999%) was evaporated at 1060 ºC, while high-purity Te (99.999%) was evaporated at 330 ºC and cracked at 1000 º C. Fe (99.99%) was evaporated following standard e-beam procedures. Fe and Te deposition rates were calibrated by AFM on pre-patterned samples. The Ge deposition rate was calibrated with homoepitaxial RHEED oscillations on Ge(111). The Fe flux rate was monitored during the deposition using a quartz microbalance, while the Ge and Te flux rates were monitored before and after the depositions with a UHV pressure gauge.

**XRD**

In-plane XRD measurements were performed with a SmartLab Rigaku diffractometer operating at 45 kV and 200 mA. Collimators with a resolution of 0.5º were used both in the source and the detector. The measurements were done with a copper rotating anode beam tube ($K_\alpha$ = 1.54 Å) at a grazing incident angle of 0.4º. The out-of-plane XRD measurements were performed using a Panalytical Empyrean diffractometer operated at 35 kV and 50 mA, with a cobalt source, ($K_\alpha$ = 1.79 Å) A PIXcel-3D detector allowed a resolution of 0.02° per pixel, in combination with a divergence slit of 0.125º.

**AFM**

The film topography was characterized with a Bruker Dimension-Icon microscope using a scanasyst-air cantilever, operated in ScanAsyst mode, and under ambient conditions. The root-mean-square roughness was evaluated on 1×1 µm$^2$ areas.

**STEM**

STEM measurements were performed using a Cs-corrected FEI Themis at 200 keV. HAADF-STEM images were acquired using a convergence semi-angle of 20 mrad and collecting scattering from >60 mrad. EDX was performed for elemental mapping using a Bruker EDX system consisting of four silicon drift detectors in the Themis microscope.

STEM specimens were prepared by the focused ion beam lift-out technique using a Zeiss DualBeam Cross Beam uSTE750 at 30 kV.

HAADF-STEM simulations were carried out using multislice simulation software, namely JEMS[37], and µSTEM[38] including the experimental conditions used for the image acquisition. The model structures were constructed based on the crystal information given in refs.[22,29].

**RBS**

Rutherford backscattering was done using an ion-beam of $^4$He$^+$ ions with 2MeV, at a chamber pressure of $10^{-6}$ mbar, with a 160º angle between the incident beam and the detector, and an angle of incidence of 75º.

**SQUID**

**S**QUID measurements were performed using a Quantum Design magnetic property measurement system MPMS®3 following standard procedures. Magnetic field sweeps were made in no-overshoot, persistent mode. The temperature sweeps were made at a rate of 5 K per minute. The estimated magnetic moments were determined from an average of two scans. For magnetization remanence measurements, the films were field-cooled from RT to 10 K with an external field of 2 T applied in-plane, and the superconducting magnet quenched before starting the temperature dependence.

**XMCD**

The XMCD measurements were performed on the DEIMOS beamline[39] of synchrotron SOLEIL. A XAS spectrum is obtained by continuously moving the energy of the monochromator over the energy range of interest while synchronizing the energy and polarization of the undulator to the energy of the monochromator[40]. The XMCD spectrum is obtained from four measurements, where both the circular helicity and the direction of the applied magnetic field are flipped.

*Ab initio* **calculations**

First-principle calculations were performed using the Vienna Ab-initio Simulation Package (VASP)[41,42], employing projector augmented wave (PAW)[43,44] for potentials with the plane wave energy cutoff in all the calculations set to 550 eV, and the generalized gradient approximation (GGA)[45] for the exchange-correlation energy. As a first step, ionic relaxation of $Fe_5GeTe_2$ bulk structure was performed with *k*-point grids of 7×7×1 and 0.01 eV/Å force tolerance during optimization of the atomic positions. The in-plane (out-of-plane) lattice parameter was fixed to 4.07 (29.3) Å. Next, a self-consistent calculation with 21×21×3 *k*-point grids was realized to obtain the charge density wave and the atomic spin moments. Finally, the orbital moments were obtained by performing a non-self-consistent calculation with spin-orbit coupling included with the quantization axis set along [001].

**Acknowledgments**

The MBE reactors were funded by the project Minatec LABS 2018 N°2018 AURA P3 and the project EPI2D of the University Grenoble Alpes IDEX. The authors acknowledge the financial support from the UGA IDEX IRS/EVASPIN, the ANR projects ELMAX (ANR-20-CE24-0015) and MAGICVALLEY (ANR-18-CE24-0007), and the King Abdullah University of Science and Technology under grant number OSR-2018-CRG7-3717, the DARPA TEE program through Grant MIPR# HR0011831554 from the DOI, and the LANEF framework (ANR-10-LABX-51-01) for its support with mutualized infrastructure.

**Author contributions:**

The sample preparation, growth, and the SQUID measurements were done by MR, GG, and FB. XRD was performed by AM. The AFM scans were made by GG. The STEM cross-sections and simulations were done by HO and DD. XMCD was done by FB, GG, DL, and PO. RBS was perform by CV, J-FJ and DJ. AH and MC performed the ab initio simulations. The study was supervised by FB, OB, and MJ. MR and FB wrote the paper, and all authors contributed to the discussion of the results.

**Competing interests:**

Authors declare that they have no competing interests.

**Data and materials availability:**

All data needed to evaluate the conclusions in the paper are present in the paper and/or the Supplementary Materials. Additional data related to this paper may be requested from the authors.

**FIGURES**

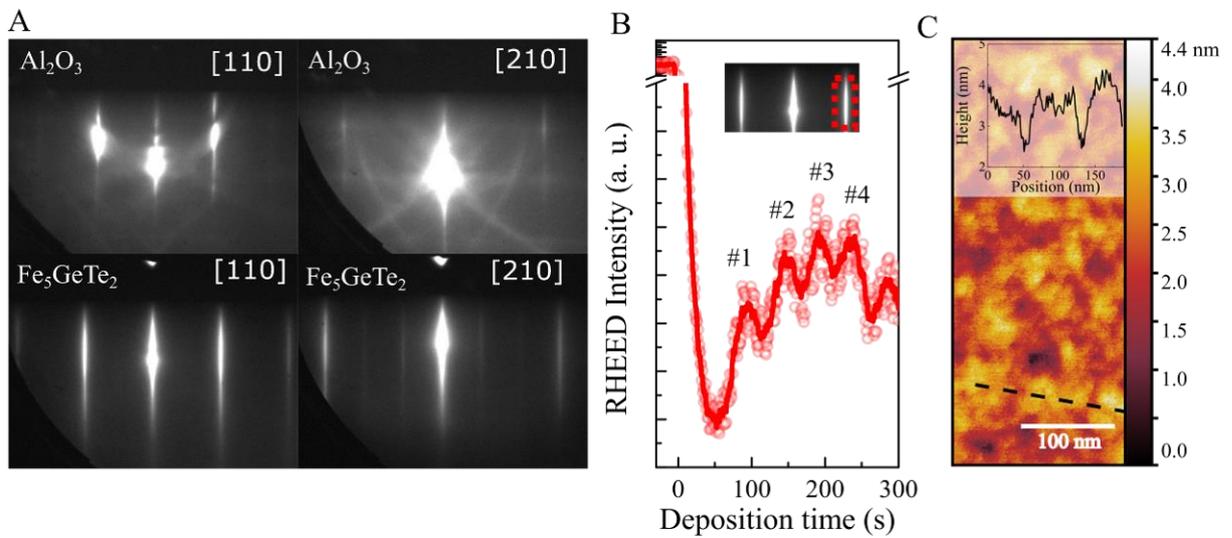

*Figure 1 **Surface characterization of Fe$_5$GeTe$_2$**. (**A**) RHEED patterns of the surface before and after the deposition of a 12-nm thick film of Fe$_5$GeTe$_2$ along the [110] and [210] azimuths. (**B**) RHEED intensity oscillations of the (010) streak. The solid line is a moving average of the raw scatter data. Monolayer number identified on each oscilation. (**C**) Atomic force microscopy image of a 12-nm-thick film. Inset: Height profile showing resolved monolayer steps (1 ML ~ 0.977 nm)*

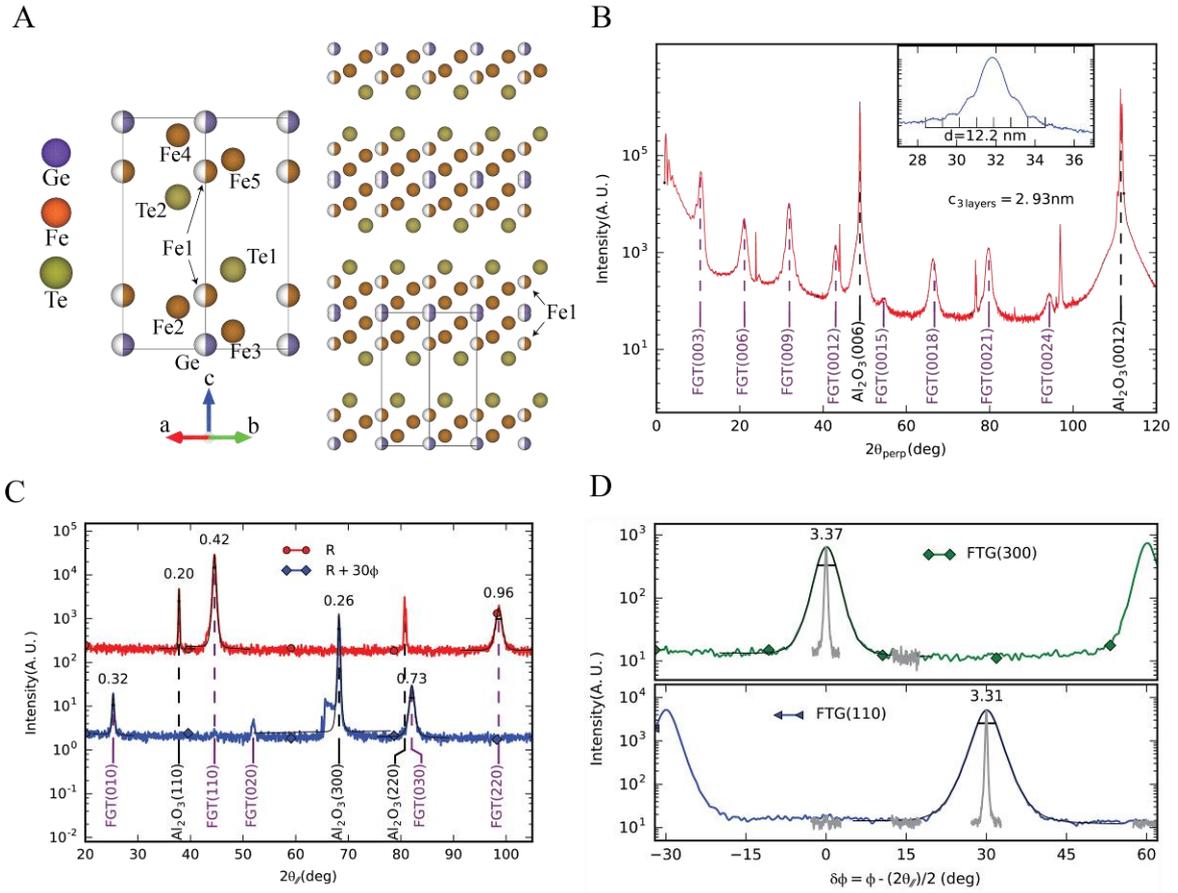

*Figure 2* **Structural characterization of $Fe_5GeTe_2$ by X-ray diffraction.** (**A**) Unit cell projected along [110], with two-halfs of an FGT layer separated by the vdW gap between adjacent Te sublayers. The right-side image shows the same projection for several repetitions of the unit cell. The occupation probability is represented by half-filled solid balls. Images generated using VESTA software[30] for the crystallographic model of ref. [22]. (**B**) Out-of-plane $\theta/2\theta$ XRD scan of the films, showing $Al_2O_3$(00l) and $Fe_5GeTe_2$ (00l) peaks. Inset: zoom on the (009) Bragg peak to show the fringes used to estimate the film thickness. (**C**) In-plane radial XRD scans along the (110) reciprocal direction (R), and (010), (R+30º). Each diffraction peak is labelled with the full widh at half maximum.(**D**) In-plane azimuthal XRD scan over the families {110} and {300} showing the 6-fold symmetry of the crystal. Grey lines show the corresponding radial peak.

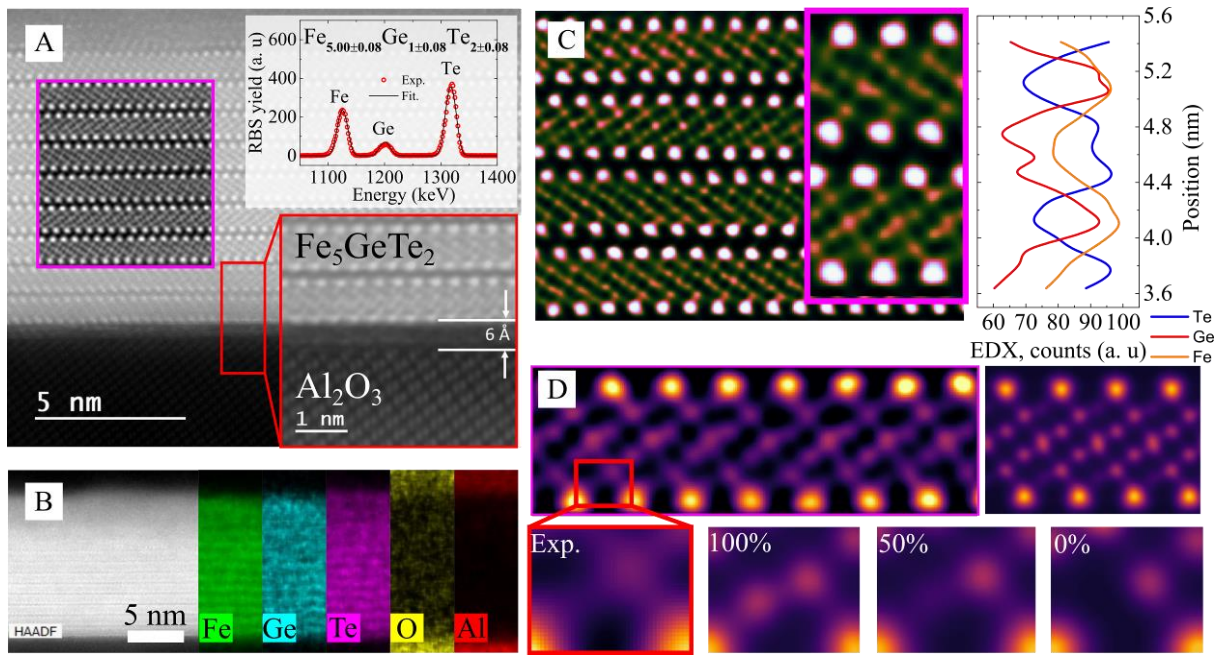

*Figure 3 **Transmission electron microscopy of Fe$_5$GeTe$_2$**. (A) High-angle annular dark-field (HAADF) cross section image of the 12 nm-thick film along [110]. Left inset: High-pass filtered image of the layers. Top inset: Rutherford backscattering spectroscopy showing stoichiometric 5:1:2 Fe:Ge:Te composition. Bottom inset: Zoomed view of the interface between Al$_2$O$_3$ and Fe$_5$GeTe$_2$. (B) Element distribution across the layers, determined using energy dispersive X-ray spectroscopy (EDX). (C) Atomic number contrast enhanced HAADF image of the Fe$_5$GeTe$_2$ layers. The left plot shows an EDX scan over two layers, selective to Te, Ge, and Fe. (D) Comparison between experimental (top left panel) and simulated (top right panel) Z-constrast enhanced HAADF images. Bottom panels are zooms on the Fe1 site. The simulations consider different occupation probabilities of the Fe1 site.*

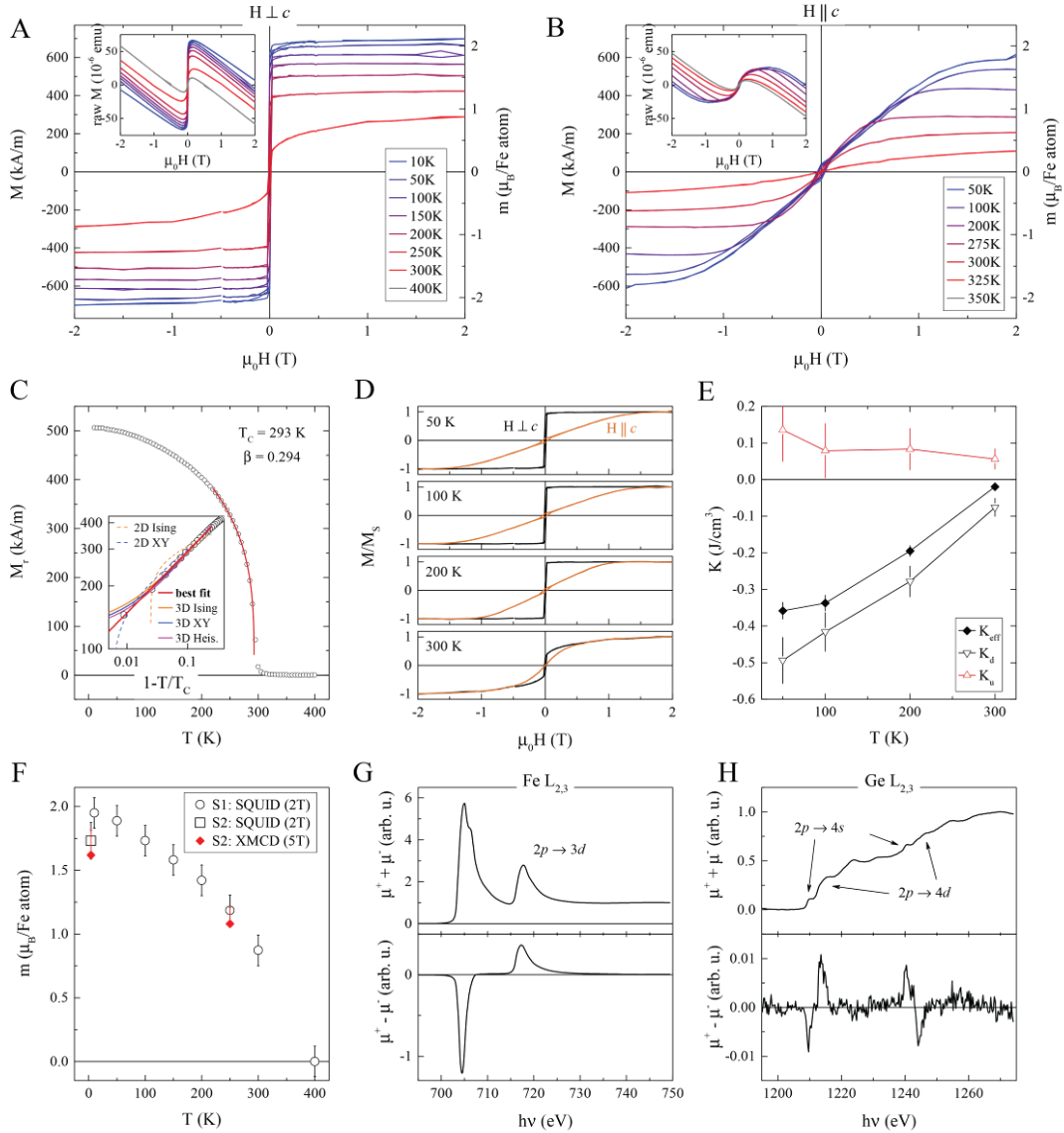

*Figure 4* **Magnetic characterization of 12-nm thick Fe$_5$GeTe$_2$ using SQUID and XMCD.** *(A) Isothermal magnetic hysteresis loops with the external magnetic field H in the ab basal plane for different temperatures. The magnetic background at temperatures well above $T_C$ has been subtracted for all measurements. Inset: Raw data highlighting the magnetic background considered. (B) Same as (A), with H parallel to the c-axis. (C) Magnetization remanence as a function of temperature. Fitting to the scaling behavior of a magnetic system in an ordered state yields $T_C ≈ 293$ K and $β≈0.294$. The inset shows the scaling behavior of the remanence for different 2D and 3D magnetic critical exponents (see Supplementary Figure 3). (D)*
24

*Normalized in-plane and out-of-plane magnetic hysteresis loops for different temperatures. (E) Effective magnetic anisotropy, $K_{eff}$, shape anisotropy, $K_d$, and uniaxial anisotropy, $K_u$, as a function of temperature. (F) Magnetic moment per Fe atom as a function of temperature determined from SQUID and XMCD for two 12-nm thick samples (S1 and S2) (G) Normalized Fe $L_{2,3}$-edge X-ray absorption spectroscopy (XAS) ($\mu^+ + \mu^-$) and XMCD ($\mu^+ - \mu^-$) spectra. $\mu^{+(-)}$ is the absorption for parallel (antiparallel) alignment of the light helicity and the sample magnetization. (H) Normalized Ge $L_{2,3}$-edge XAS and XMCD.*

**TABLES**

Table 1 Spin ($m_S$), orbital ($m_L$) and total ($m_S+m_L$) magnetic moment per atom in $Fe_5GeTe_2$, estimated from first principles calculations. A single Fe1 site was considered to have 100% occupancy (see text). The last two lines give the calculated averaged Fe magnetic moments and the experimental values measured by XMCD, respectively.

| Atom | $m_S$ ($\mu_B$/at.) | $m_L$ ($\mu_B$/at.) | $m_S + m_L$ ($\mu_B$/at.) |
|---|---|---|---|
| Te1 | 0.00 | 0.00 | 0.00 |
| Fe1 | -0.16 | -0.03 | -0.19 |
| Fe2 | 2.35 | 0.05 | 2.40 |
| Fe3 | 2.10 | 0.04 | 2.14 |
| Ge | -0.06 | 0.00 | -0.06 |
| Fe4 | 1.50 | 0.03 | 1.53 |
| Fe5 | 2.60 | 0.08 | 2.68 |
| Te2 | -0.10 | -0.02 | -0.12 |
| Fe (averaged) | 1.68 | 0.03 | 1.71 |
| Fe (exp.) | 1.54 | 0.08 | 1.62 |